\documentclass[prb,aps,amsmath,amssymb,reprint]{revtex4-1}
\usepackage{graphicx}
\usepackage{dcolumn}
\usepackage{bm}
\begin{document}
\title{Surfactant induced smooth and symmetric interfaces in Cu/Co multilayers}
\author{S. M. Amir, Mukul Gupta}
\email[]{mgupta@csr.res.in}
\author{Ajay Gupta}
\affiliation{UGC-DAE Consortium for Scientific Research,
\\University Campus, Khandwa Road, Indore-452 001,India}
\author{J. Stahn}
\affiliation{Laboratory for Neutron Scattering, ETH Zurich and
Paul Scherrer Institut, CH-5232 Villigen PSI, Switzerland}
\author{A. Wildes}
\affiliation{Institut Laue-Langevin, rue des Martyrs, 38042
Grenoble Cedex, France}

\date{\today}

\begin{abstract}
In this work we studied Ag surfactant induced growth of Cu/Co
multilayers. The Cu/Co multilayers were deposited using Ag
surfactant by ion beam sputtering technique. It was found that Ag
surfactant balances the asymmetry between the surface free energy
of Cu and Co. As a result, the Co-on-Cu and Cu-on-Co interfaces
become sharp and symmetric and thereby improve the thermal
stability of the multilayer. On the basis of obtained results, a
mechanism leading to symmetric and stable interfaces in Cu/Co
multilayers is discussed.
\end{abstract}

\pacs{74.78.Fk, 61.05.fj , 61.05.cm, 61.05.cp, 75.70.-i}
\keywords{thin film, sputtering, evaporation, neutron
reflectivity, x-ray reflectivity, magnetoresistance}

\maketitle

\section{Introduction}
Smooth and symmetric interfaces of magnetic multilayers are of
much interest because of their application in technological
devices like recording media, read heads and
sensors.~\cite{Wolf_Science2001,Prinz_Science98,carcia_JAP1988}
The magnetic layers separated by the non magnetic spacer layer
show the giant magnetoresistance (GMR) due to oscillatory exchange
coupling for varying spacer layer
thicknesses.~\cite{Parkin_PRL1990,Parkin_PRL1991} Thickness
fluctuation of the spacer layer due to the interface roughness
affect the strength of exchange coupling.~\cite{Schmidt_PRB1999}
GMR largely depends on the asymmetry of the roughness at the
interface due to the spin dependent scattering across the
interfaces.~\cite{Parkin_PRL1993,Zahn_PRL98,Kief_PRB93} It was
experimentally evidenced that sputtered Cu/Co multilayer exhibit
the largest GMR with oscillatory exchange
coupling.~\cite{Parkin_PRL1991} The GMR effect was first observed
in Fe/Cr superlattices~\cite{Baibich:PRL88} and experimental and
theoretical studies demonstrate that GMR in Fe/Cr
multilayer~\cite{Fullerton:PRL92} increases with the presence of
roughness at magnetic/nonmagnetic interface while it decreases
with interface roughness for Cu/Co multilayer.~\cite{Hall:PRB93}
Therefore smooth and symmetric interfaces in Cu/Co multilayer are
essential for application point of view.

Asymmetric interface occurs in multilayers due to the difference
in the surface free energy ($\gamma$) of the
elements.~\cite{egelhoff_JVT1989} In Cu/Co multilayer system
$\gamma$ for Cu and Co is different. As such the surface free
energy depends on the crystallographic orientation, in case of
polycrystalline structures, $\gamma$ for average face is relevant.
The experimentally observed values of $\gamma$ for the average
face of Cu, Co and Ag are 1.8\,Jm$^{-2}$, 2.55\,Jm$^{-2}$ and
1.24\,Jm$^{-2}$, respectively.~\cite{Foiles_PRB86,Tyson_SS77} This
difference in the surface free energy leads to wetting of Cu-on-Co
and de-wetting of Co-on-Cu, giving rise to a smooth Cu-on-Co and a
rough Co-on-Cu interface. This was experimentally evidenced by
Timothy \textit{et al.} in scanning tunnelling microscopy
demonstrating that Co make islands over Cu while Cu make a smooth
layer on Co in Cu/Co multilayers.~\cite{Timothy:JAP96}

It has been demonstrated in the literature that by using a surface
active species or so called \textit{surfactant} the difference
between the surface free energies can be minimized. Use of such
surfactants in the crystal growth technology is well
established.~\cite{Buckley} However, in case of thin films
deposited in vacuum, surfactants have been used only in few
multilayer
systems.~\cite{Copel_PRL89,Egelhoff_JAP1996,Egelhoff_JAP96_Pb_Au,Egelhof_JAP97_O,osten_APL92,Camarero_PRL94,Camarero_PRL96,Chopra_PRB97}
Egellhoff and Steigerwald~\cite{egelhoff_JVT1989} studied the role
of adsorbed gases (H, O, N, Co and S) in deposition of
metal-on-metal epitaxial systems. These adsorbed gases float or
segregate to the surface, balancing the surface/interface energy
and strain during the growth. Gaseous surfactants such as oxygen,
also suppresses the intermixing and increases the GMR by
restricting pinholes in Cu spacer layer during
deposition.~\cite{Tolkes_PRL98,Wen_PRB2007,Larson_PRB2003}
Although, due to enhanced mobility of gaseous surfactants, it is
likely that they get trapped across the grain boundaries.

In case of Ge/Si(100) multilayers, Copel \textit{et al.} have
demonstrated that use of As surfactant triggers the layer-by-layer
type growth and inhibits interdiffusion.~\cite{Copel_PRL89} Hoegen
\textit{et al.} have shown that Sb surfactant not only inhibits
interdiffusion but results in a relaxed (strain free) and defect
free Ge film on Si(111).~\cite{Hoegen_PRL91} In a theoretical
study by Barab$\acute{\mathrm{a}}$si the interaction of surface
with a surfactant was described.~\cite{Barabasi_PRL1993} In an
another theoretical work by Zhang and Lagally~\cite{Zhang_PRL94}
the surfactant mediated layer-by-layer growth was described on the
basis of atomic interactions. Recently, Egelhoff and co-workers
have explored the effect of surfactant (e.g In, Ag, O, Pb etc.) in
Cu based spin valve systems and demonstrated that surfactant
improves the surface and interface property as well as increases
the GMR value.~\cite{Egelhoff_JAP1996,yang_JAP2001,Chopra:PRB2002}
In an another study, Camarero \textit{et al.} have demonstrated
that Pb atoms used as surfactant suppresses the twin formation
which increases the coupling between Co
layers.~\cite{Camarero_PRL94} Theoretical studies also show that
monolayer of Pb used as a surfactant in Cu/Co multilayer minimizes
the difference in surface free energy of Cu and Co, inhibits the
island formation and floats over the surface by atomic exchange
process.~\cite{Gomez:PRB01,kim:JAP2009} The use of Ag surfactant
was also studied to examine the interfacial
intermixing~\cite{an_JAP2006,Peterson:PhysicaB2003} and GMR in
magnetic multilayers.~\cite{Zou:PRB2001} In case of Ti/Ni
multilayers Ag surfactant was used to get smooth and symmetric
interface.~\cite{TiNi_arXiv} However, detailed studies on the Ag
surfactant mediated growth in Cu/Co multilayers have not been
performed.

In the present work we studied the effect of Ag surfactant in
Cu/Co multilayer. It was also investigated how this addition of Ag
surfactant affects the structural and magnetic properties of Cu/Co
multilayers. We used neutron reflectivity (NR) technique which is
non destructive and measures the interface roughness with an
accuracy less than an \AA. In addition, we used NR to measure the
interdiffusion across Cu/Co interfaces. It was found that addition
of Ag surfactant makes Cu-on-Co and Co-on-Cu interfaces smooth and
symmetric which are otherwise rough and asymmetric. This leads to
reduced interdiffusion and thereby improved thermal stability of
Cu/Co multilayers prepared using Ag surfactant. The obtained
results are presented and discussed in the following sections.

\section{Experimental Details}

Ion beam sputtering (IBS) technique was used to deposit the Cu/Co
multilayers with and without Ag surfactant. The Ar$^+$ ions of
energy 1\,keV were produced using a radio-frequency ion beam
source (Veeco 3cm RF source). The ion beam was neutralized using a
RF generated electron flood source. The ion beam of size 3\,cm was
kept incident at an angle of 45\,$^{\circ}$ with respect to a
target. The targets were mounted on a rotary motion feedthrough
which can hold up to four different targets. The targets were
sputtered alternatively to deposit a multilayer structure. The
samples were prepared without any surfactant as a reference
(sample $A$) and with Ag surfactant (sample $B$) added on top of a
Cu buffer layer deposited on a Si(100) substrate at room
temperature (without intentional heating). The nominal structure
of samples are given below:

($A$) Cu\,(10\,nm)/[Cu\,(3\,nm)/Co\,(2\,nm)]$_{10}$

($B$) Cu\,(10\,nm)/Ag\,(0.2\,nm)/[Cu\,(3\,nm)/Co\,(2\,nm)]$_{10}$

Here, the Cu layer thickness of 3\,nm corresponds to the third
(and weakest) AF peak in the oscillatory exchange coupling. At
this thickness it is expected that magnetoresistance (MR) will be
small compared to the first or second maxima around 1\,nm and
2\,nm, respectively.~\cite{Parkin_PRL1991} The Cu layer thickness
of 3\,nm was chosen as at lower thickness the Bragg peak in
neutron reflectivity (NR) will appear at high q$_z$ values and due
to limited flux of neutron sources it may be difficult to do NR
measurements in a reasonable time.

Prior to the deposition of samples the base pressure was about
$2\cdot10^{-8}$ mbar and during deposition the pressure was about
$5\cdot10^{-4}$ mbar due to flow of Ar gas (purity 99.9995\%) in
the source and neutralizer. X-ray and neutron reflectivity
measurements were carried out to measure the thickness, surface
and interface roughness of the samples. X-ray reflectivity (XRR)
measurements were carried out using x-rays of wavelength
1.54\,\AA~ generated using a laboratory source. The NR
measurements were carried out at the SuperADAM instrument at ILL,
Grenoble, France using neutrons of wavelength 4.4\,\AA. In order
to study the the thermal stability of the samples, the NR
measurements were also carried out at NARZISS reflectometer at
SINQ/PSI. X-ray diffraction (XRD) measurements were carried out
with 1.54\,\AA~ x-rays in the $\theta$-$2~\theta$ geometry using a
standard diffractometer (Bruker D8 Advance) equipped with a fast
1-D detector based on silicon drift technology (Bruker LynxEye).

The magnetoresistance (MR) measurement for all the samples were
carried out at room temperatures using four point probe method.
The direction of the current flowing in the sample was along the
direction of the magnetic field (parallel to the surface of the
sample). Here, the MR is defined as MR = ($R_{\circ}$ -
$R_{sat}$)/$R_{\circ}$, where $R_{\circ}$ is the resistance in the
absence of magnetic field and $R_{sat}$ is the resistance under
the magnetic field in which the sample are magnetically saturated.

\section{Results}

\subsection{X-ray and neutron reflectivity measurement}

\begin{figure}
\includegraphics[width=80mm,height=80mm]{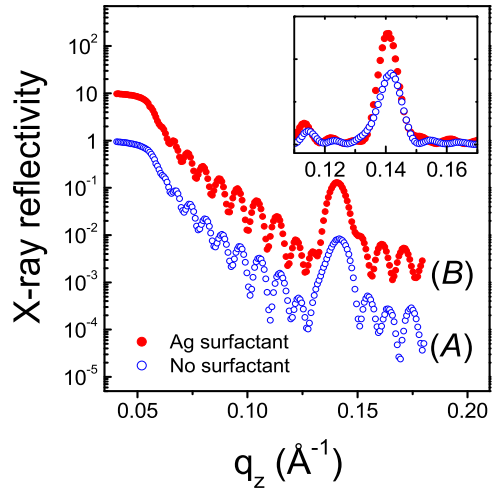} \vspace{-5mm}
\caption{\label{fig:1} (color online) X-ray reflectivity of Cu/Co
multilayers deposited without surfactant ($A$) with Ag surfactant
($B$. Inset show the Bragg peak intensity at Bragg peak.The
pattern on y-axis have been shifted for clarity. \vspace{-1mm}}
\end{figure}

\begin{figure} \vspace{-5mm}
\includegraphics [width=90mm,height=110mm] {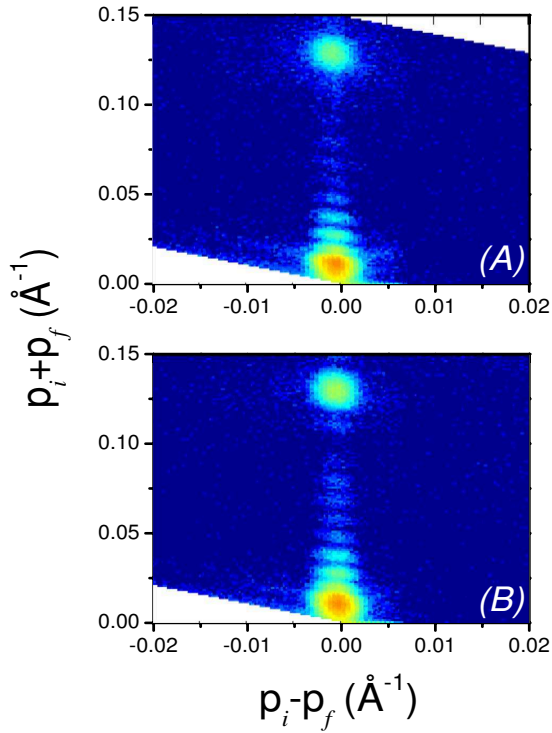}
\vspace{-5mm} \caption{\label{fig:2} (color online) 2-D Position
sensitive detector (PSD) image of Cu/Co multilayers prepared
without surfactant ($A$) and with Ag surfactant ($B$).
\vspace{-5mm}}
\end{figure}

\begin{figure}
\includegraphics [width=80mm,height=80mm] {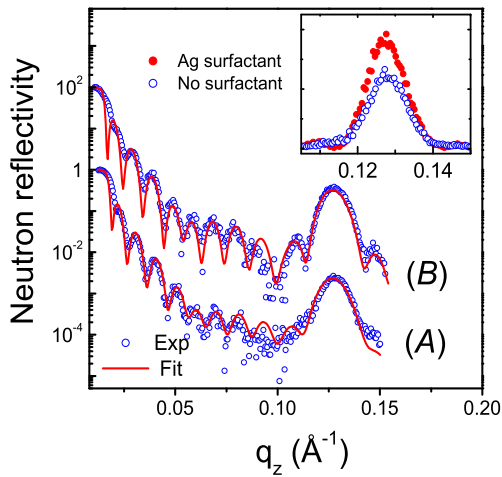} \vspace{-7mm}
\caption{\label{fig:3} (color online) Neutron reflectivity pattern
Cu/Co multilayers prepared without surfactant ($A$) and with Ag
surfactant ($B$). Inset shows the reflectivity at Bragg peak. The
pattern on y-axis have been shifted for clarity. \vspace{-7mm}}
\end{figure}

\begin{figure}
\includegraphics[width=80mm,height=70mm]{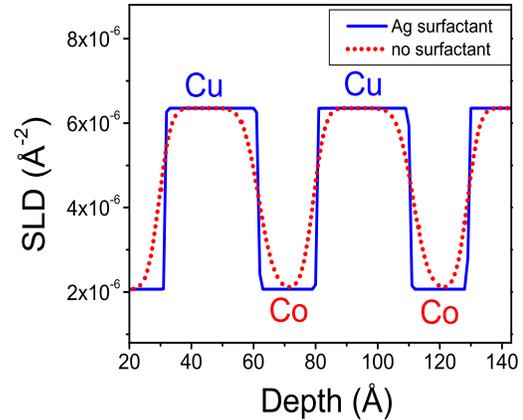} \caption{\label{fig:4}(color online) Scattering
length density profile of with surfactant (solid line) and without
Ag surfactant (dash line) samples of Cu/Co multilayer.}
\end{figure}

Fig.~\ref{fig:1} shows the x-ray reflectivity pattern of the
as-deposited samples. The XRR pattern for the samples $A$ is shown
in (fig.~\ref{fig:1}\,$A$) and for the sample $B$ in
(fig.~\ref{fig:1}\,$B$). As can be seen from the figure, the XRR
pattern shows a Bragg peak around q$_{z}$=0.14\,\AA$^{-1}$,
corresponding to the bilayer period of 4.9\,nm which is close to
the nominal value of 5\,nm. The oscillation in the pattern
correspond to the total thickness of the sample. As can be seen
from the inset of fig.~\ref{fig:1}, the intensity at the position
of Bragg peak is enhanced significantly when Ag surfactant was
added to the multilayer structure. Such an enhancement in the
intensity of the Bragg peak indicates that the interface get
smoother with addition of Ag surfactant. It may be noted that the
contrast between Cu and Co for the x-rays of wavelength used in
this work is rather poor due to a small difference between the
electron density of Cu and Co. For x-rays the refractive index is
defined as $n=1-\delta-i\beta$ and with 1.54\,\AA~x-rays the
dispersion ($\delta$) and absorption ($\beta$) parts of optical
constants in terms of number density are: $\delta_{\mathrm{Cu}}$ =
6.45$ \cdot 10^{-5}$\,\AA$^{-2}$, $\beta_{\mathrm{Cu}}$ = 1.45$
\cdot 10^{-6}$\,\AA$^{-2}$, $\delta_{\mathrm{Co}}$ = 6.30$ \cdot
10^{-5}$\,\AA$^{-2}$, $\beta_{\mathrm{Cu}}$ = 9.135$ \cdot
10^{-6}$\,\AA$^{-2}$. Therefore the Bragg peak appearing in
fig.~\ref{fig:1} is basically due to contrast in the absorption
part of the refractive index which makes it rather difficult to
model the x-ray reflectivity data. However, qualitatively it can
be seen that addition of Ag surfactant helps in reducing the
interface roughness. Therefore, in order to get more insight into
the interfaces of the multilayers, we performed neutron
reflectivity measurements. In case of neutrons the contrast
between Cu and Co is much larger as compared to x-rays, with
absorption being negligible due to much weaker interaction of
neutrons with matter as compared to x-rays. The values of
scattering length density (SLD) for Cu and Co in case of neutrons
are 6.55$\cdot$10$^{-6}$\,\AA$^{-2}$ and
2.26$\cdot$10$^{-6}$\,\AA$^{-2}$, respectively.

The neutron reflectivity pattern of samples were recorded using a
2-D position sensitive detector (PSD). The PSD images of samples
are shown in fig.~\ref{fig:2} as a function of
($\mathrm{p}_i+\mathrm{p}_f$) and ($\mathrm{p}_i-\mathrm{p}_f$)
with $\mathrm{p}_{i(f)} = 2\pi~\mathrm{sin~}\theta_{
i(f)}/\lambda$ the normal to surface component of the incoming
(outgoing) wave vector and $\theta_{i(f)}$ the angle of the
incidence (scattering) and $\lambda$ being the wavelength of the
neutrons.~\cite{pi-pf_PRL_2002} A cut across the line for
$\mathrm{p}_{i} = \mathrm{p}_{f}$ gives the specular reflectivity
which is plotted in fig.~\ref{fig:3} for samples $A$ and $B$ as a
function of $q_z$ ($\mathrm{q}_z = \mathrm{p}_{i} +
\mathrm{p}_{f}$). Here it may be noted that in the NR pattern we
did not observe magnetic Bragg peak which is expected due to
antiferromagnetic (AF) coupling.~\cite{Langridge:PRL2000} This may
be due to the fact that the thickness of the Cu layer correspond
to the third AF peak which is weakest in the oscillatory exchange
coupling. Further since we have only 10 repeats of Cu/Co bilayers,
it is expected that the intensity of the AF peak will be too small
to measure experimentally in this case. This is further confirmed
by the MR measurements (shown later) where the typical values of
MR are about 1\% in the as-deposited samples.

As can be seen from the PSD images, the Bragg peak appears as a
bright spot  for both samples. The brightness of the Bragg
reflected region is more intense for sample $B$ as compared to
sample $A$. This shows that with addition of Ag surfactant the
reflectivity at the Bragg peak enhances as also observed with
x-ray reflectivity data. Such an enhancement in the Bragg peak may
happen due to a reduction in the interface roughness. If the
roughness decreases, it should result in less diffuse scattering
which appears along the x-direction ($\mathrm{p}_i-\mathrm{p}_f$)
in fig.~\ref{fig:2}. A closer look at the PSD images indeed shows
lesser diffuse scattering for sample $B$ around the critical edge
and Bragg peak positions. However, as shown later, the roughnesses
are rather small therefore, the diffuse scattering expected from
these multilayers is small.

Fig.~\ref{fig:3} shows the deduced NR pattern for samples $A$ and
$B$. The patterns were fitted using Parratt's formalism
~\cite{Parratt32} and the fitted results are given in
table~\ref{tab:table1}. As can be seen from the table, the value
of reflectivity at the Bragg peak increases from 0.23\% to 0.36\%
when Ag surfactant was added. The roughness of Co-on-Cu interface
and Cu-on-Co interface was 0.38\,nm and 0.18\,nm, respectively
when no surfactant was used. With addition of Ag surfactant the
interface roughness of both interface becomes equal at about
0.1\,nm. This value of roughness appears rather small, which may
be due to the fact that we did not subtract the diffuse
scattering. Therefore the value of roughnesses should be taken as
a lower limit. Fig.~\ref{fig:4} shows the SLD profile of samples
$A$ and $B$, obtained from the fitting of NR data. As can be seen
from the figure, the SLD profiles were asymmetric and broad when
no surfactant was used, with addition of Ag surfactant the
profiles become sharp and symmetric. This clearly shows that the
addition of Ag surfactant in the Cu/Co multilayer results in an
appreciable decrease of interface roughnesses and also the of
Co-on-Cu and Cu-on-Co interfaces become symmetric.

\begin{table}
\caption{\label{tab:table1} Intensity at Bragg peak in Neutron
reflectivity measurement of with and without Ag samples deposited
by ion beam sputtering.}
\begin{ruledtabular}
\begin{tabular}{c|c|c}
Sample& No surfactant & Ag surfactant\\ \hline
R$_{\mathrm{Bragg}}$(\%)& 0.23$\pm0.01$ & 0.36$\pm0.01$ \\
d (nm) & 4.9$\pm0.1$ & 4.9$\pm0.1$ \\
$\sigma$$_{\mathrm{[Co-on-Cu]}}$ (nm)& 0.36$\pm0.01$  & 0.11$\pm0.01$\\
$\sigma$$_{\mathrm{[Cu-on-Cuo]}}$ (nm)& 0.18$\pm0.01$  & 0.10$\pm0.01$\\
\end{tabular}
\end{ruledtabular}
\end{table}

\subsection{Thermal stability of Cu/Co multilayer}

The thermal stability of the Cu/Co multilayer was studied by doing
neutron reflectivity, x-ray diffraction and magnetoresistance
measurements in the samples $A$ and $B$ after annealing the
samples in a vacuum furnace with a base pressure of about
1$\cdot$10$^{-6}$\,mbar. The annealing of the samples was
performed between 373\,K to 673\,K with a step of 100\,K. It was
found that up to a temperature of 473\,K the properties of samples
remain identical to the as-deposited samples, however above this
temperature, the samples prepared using Ag surfactant were found
more stable as compared to samples prepared without any
surfactant. The results presented in the following subsections:

\subsubsection {Neutron reflectivity measurements}

\begin{figure}
\includegraphics[width=80mm,height=120mm]{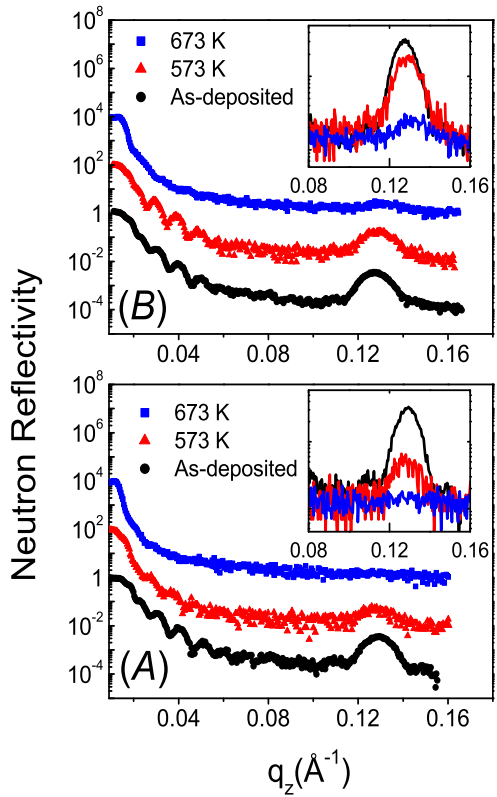} \caption{\label{fig:5} (color online) Neutron
reflectivity of Cu/Co multilayers prepared without any surfactant
($A$) and with Ag surfactant ($B$) after annealing at different
annealing temperatures. The pattern were shifted on y-scale for
clarity. Inset shows a comparison of the intensity at the Bragg
peak.}
\end{figure}

In order to check the thermal stability, the NR measurements were
performed at the NARZISS reflectometer at SINQ/PSI. The neutron
reflectivity pattern of the samples prepared with and without Ag
as a surfactant are shown in fig~\ref{fig:5}. The measurements
shown are in the as-deposited samples and after annealing the
samples at 573\,K, and at 673\,K for 1\,hour at each temperature.
The Bragg peak corresponding to the multilayer period appears at
nearly the same angles in both samples. The intensity of the Bragg
peak is expected to decay due to the inter-diffusion as the
annealing temperature is raised. This decay is seen to be faster
in the samples when no surfactant was used.

In the sample $B$, the Bragg peak intensity decreases marginally
at temperature of 573\,K. However, in case of sample $A$ where no
Ag surfactant was used at 573\,K there is significant decrease of
Bragg peak intensity. At 673\,K the Bragg peak completely
disappears for the sample $A$ while for the sample $B$ the Bragg
peak can still be seen, however its intensity is also
significantly reduced. This result clearly demonstrates that the
thermal stability of the Cu/Co multilayer is improved with Ag
surfactant. From the measured neutron reflectivity data the
inter-diffusion in both samples can be quantified and the decay of
the Bragg peak intensity can be used to calculate the diffusion
coefficient using the
expression.~\cite{MGupta_PRB2004,rosenblum_APL1980}

\begin{equation}
\label{eq:1} I(t) = I(0)\, \exp \left(-\frac{8 \pi^{2}\, D}
{\ell^{2}}\,t\right),
\end{equation}

\begin{table}
\caption{\label{tab:table2} Inter diffusion length (L$_d$) and
diffusivity (D) obtained from Neutron reflectivity measurement of
with and without Ag samples.}
\begin{ruledtabular}
\begin{tabular}{l|ll|ll}
Sample& No surfactant & & Ag surfactant&\\ \hline

&D(m$^2s^{-1}$)& L$_d$ (nm) &D (m$^2s^{-1}$)&L$_d$ (nm)\\ \hline

573\,K& 1.9($\pm$0.3$)\cdot10^{-22}$ & 2($\pm0.2$)&5.1($\pm0.5$)$\cdot10^{-23}$&1.0($\pm0.2$)\\
\hline

673\,K& -- & -- & 2.8($\pm0.3$)$\cdot10^{-22}$&2.5($\pm0.3$) \\

\end{tabular}
\end{ruledtabular}
\end{table}

where $I(0)$ is the intensity before annealing and $I(t)$ is the
intensity after annealing time $t$ at temperature T, $\ell$ is the
bilayer periodicity. With known diffusion coefficient ($D$)
calculated using eq.~\ref{eq:1} the inter diffusion length
($L_{d}$) can be calculated with the expression $L_{d}^{2}=6Dt$ in
the direction normal to the surface of
samples.~\cite{Schmidt:AM:2008} The inter-diffusion lengths
obtained in this way are given in table~\ref{tab:table2} along
with the values of diffusion coefficient $D$. Clearly, the
inter-diffusion length $L_{d}$ is significantly smaller for the
sample prepared with Ag surfactant as compared to the sample
prepared without any surfactant. Therefore, by using Ag surfactant
the inter-diffusion and diffusivity in Cu/Co multilayers is
reduced significantly.

\subsubsection {X-ray diffraction measurements}
\begin{figure}
\includegraphics[width=90mm,height=65mm]{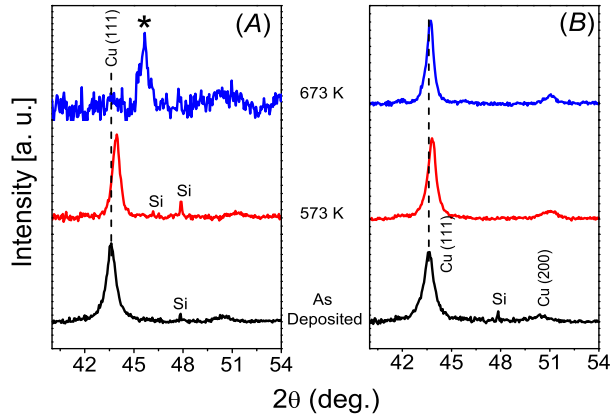} \caption{\label{fig:6} (color online) XRD
pattern of Cu/Co multilayers prepared without surfactant ($A$) and
using Ag surfactant ($B$) at different annealing temperatures.}
\end{figure}

The XRD pattern of Cu/Co multilayers prepared with and without Ag
surfactant in the as-deposited state and after annealing at
temperatures of 573\,K and 673\,K are shown in fig.~\ref{fig:6}.
The XRD pattern of the as-deposited samples show a sharp peak
around 2$\theta$ $=$ 43.6$^{\circ}$ and a broader peak around
2$\theta$ $=$50.5$^{\circ}$ corresponding to Cu\,(111) and
Cu\,(200) reflections, respectively. After annealing at 573\,K,
these peaks shift towards the higher angle side both in sample $A$
and $B$, which indicates a reduction in the inter-atomic distance.
Further annealing at 673\,K results in complete suppression of
Cu\,(111) peak for the sample prepared without any surfactant,
whereas in the sample prepared using Ag surfactant this peak
remains intact. It is interesting to see that in the sample
prepared without any surfactant a new peak appears at 2$\theta$
$=$45.6$^{\circ}$ (shown as star in fig.~\ref{fig:6}), which does
not correspond to any known phases of Cu or Co. As Cu-Co is an
immiscible system, therefore this peak can not be assigned to a
known alloy of CuCo. In order to confirm our results, we repeated
this experiment by preparing a new Cu/Co multilayer sample both
with and without surfactant and observed the same results as shown
in fig.~\ref{fig:6}. It has been reported in the literature that
an intermixed fcc-structured superlattice phase of CuCo may form
upon solid-state interfacial reaction after annealing at moderate
temperature around 573\,K.~\cite{Li:PRB01} However, more details
about such superlattice structure are not available.

From the interdiffusion measurements using neutron reflectivity as
we observed that the interdiffusion in the sample prepared without
any surfactant was significantly enhanced as compared to the
sample prepared using Ag surfactant. An enhanced interdiffusion
may give rise to an intermixed CuCo phase as observed in the
present case. However, at 673\,K the interdiffusion even in the
sample prepared with surfactant is similar to the sample prepared
without surfactant at 573\,K, still no such intermixed CuCo phase
can be observed in the sample prepared using Ag surfactant. As
pointed out by Li \textit{et al.},~\cite{Li:PRB01} the origin of
intermixed CuCo phase may be due to the excess surface free
energy. In the case when a sample is prepared using Ag surfactant
the surface free energy of Cu and Co are balanced and in this
situation the excess free energy for formation of an intermixed
CuCo phase may not be available. Therefore our results clearly
demonstrate that by balancing the surface free energy using Ag
surfactant the thermal stability of Cu/Co multilayers can be
enhanced.

\begin{figure}
\includegraphics[width=90mm,height=65mm]{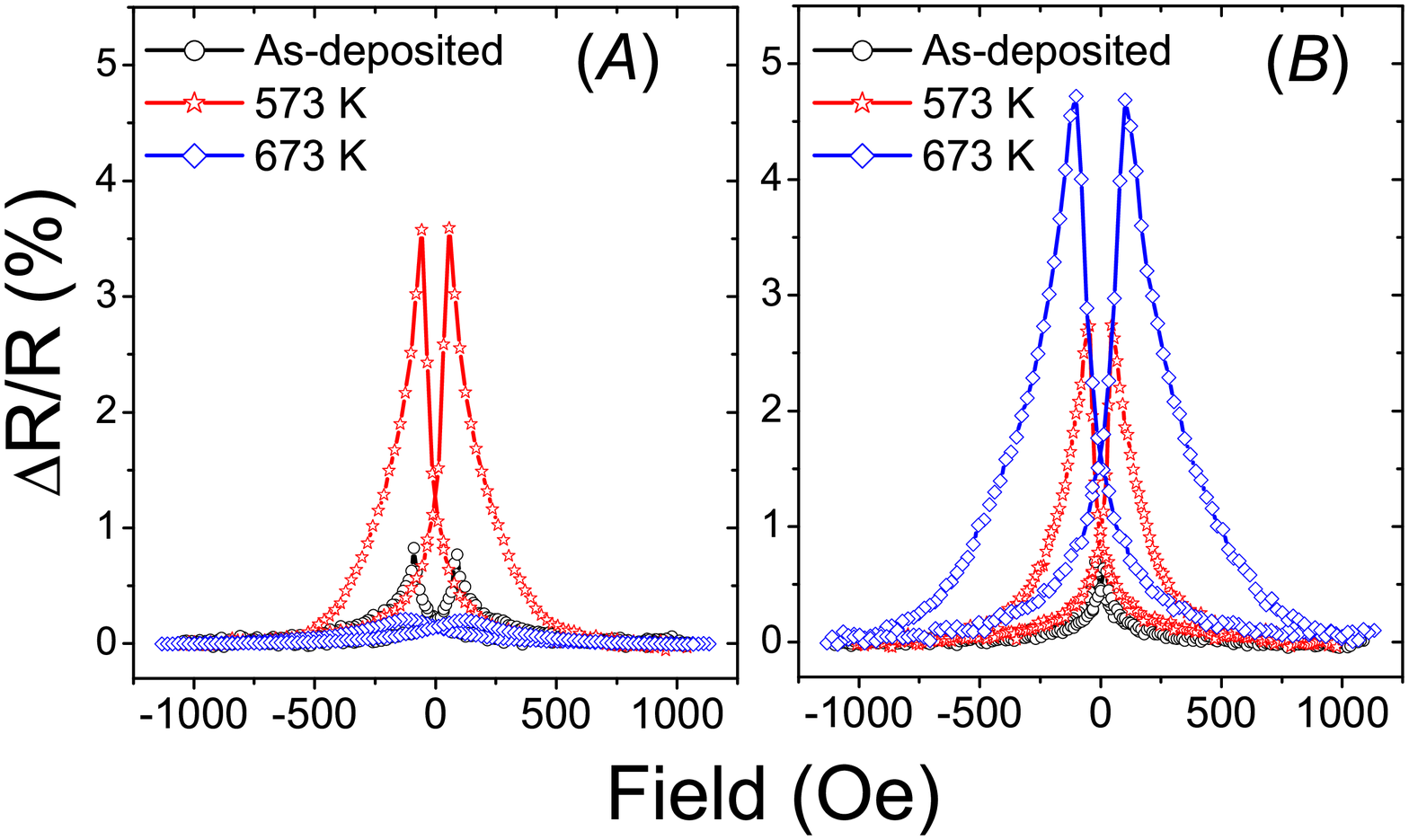} \caption{\label{fig:7} (color online) MR of
sample $A$ and sample $B$ at different annealing temperature.}
\end{figure}

\subsubsection {Magnetoresistance measurements}

Fig.~\ref{fig:7} shows the MR of Cu/Co multilayers prepared
without surfactant (fig.~\ref{fig:7}\,$A$) and with Ag surfactant
(fig.~\ref{fig:7}\,$B$). In case of sample $A$, the value of MR
increases after annealing at 573\,K and suppresses completely
after annealing at 673\,K. Whereas in case of sample $B$, the
value of MR increases both at 573\,K and 673\,K. The obtained MR
results can be understood in correlation with NR and XRD results.
As can be seen from the NR and XRD measurements, for the sample
$A$, the Cu and Co layer interdiffuse appreciably at 573\,K.
Therefore an increase in the MR at 573\,K may be due to formation
of a granular structure as a result of interdiffusion of Cu and
Co. In this situation the distance between Co atoms separated Cu
atoms may get decreased which may result in an exchange coupling
between Co atoms. However, at 673\,K when an intermixed CuCo phase
is formed, the MR suppresses completely due to formation of this
intermixed phase. In the sample prepared using Ag surfactant since
no such intermixed phase is formed, MR still increases at 673\,K.

\section{Discussions}
From the results presented in the previous sections, the effect of
Ag surfactant in the Cu/Co multilayer may be summarized as: (i)
reduction of interface roughness leading to symmetry of Co-on-Cu
and Cu-on-Co interfaces (ii) suppression of interdiffusion and
(iii) improvement in the thermal stability of the multilayer.

As mentioned before $\gamma_{\mathrm{Cu}}$=1.8\,J/m$^2$ and
$\gamma_{\mathrm{Co}}$=2.55\,J/m$^2$. This difference in the
surface free energy leads to asymmetric Co-on-Cu and Cu-on-Co
interfaces. When Cu gets deposited on Co, it will wet the surface
of Co as the surface free energy of Cu is smaller. On the other
hand when Co is getting deposited on Cu, it will de-wet or
agglomerate on Cu. This situation will lead to a sharper Cu-on-Co
interface as compared to Co-on-Cu interface. Ag surfactant with
$\gamma_{\mathrm{Ag}}$=1.24\,J/m$^2$, will help in removing the
asymmetry due to its very low surface free energy as compared Cu
and Co. In this condition when the growth of the multilayer takes
place, the upcoming atoms basically see the lower surface free
energy of surfactant atoms instead of other element of the
multilayer. This leads to wetting of the upcoming layer. If the
surfactant atoms float off to the surface, the deposition of next
element will also see the surface free energy of the surfactant.
In this situation the layer-by-layer type growth is induced
resulting in smooth and symmetric interface in a
multilayer.~\cite{Copel_PRL89,Barabasi_PRL1993}

In our case the obtained results indicate that addition of Ag
surfactant is altering the growth mode of the Cu/Co multilayer.
The surfactant atoms placed once on the Cu buffer layer balance
the surface free energy of Cu and Co resulting in symmetric
Cu-on-Co and Co-on-Cu interfaces. As expected, in absence of Ag
surfactant, the interface roughness of Co-on-Cu interface is
larger as compared to Cu-on-Co interface. An asymmetry in the
interface roughness may result in strained interfacial
region~\cite{Hoegen_PRL91} which will act like nucleation centers
when the multilayer is annealed at higher temperatures. This will
result in enhanced interdiffusion as observed in our case. Whereas
since the addition of Ag surfactant removes the asymmetry of the
interfaces, such strained regions may be minimized to a large
extent resulting in stable interfaces. The reduction of
interdiffusion length may be therefore understood in terms of
smooth and sharper interfaces formed by addition of Ag surfactant.
The XRD results obtained in this work also support this argument
as in absence of Ag surfactant a superlattice Cu/Co
structure~\cite{Li:PRB01} is formed while no such structure can be
observed when Ag surfactant was added.

\section{Conclusions}

Therefore in conclusion from the present study we show that
addition of Ag surfactant results in smooth and symmetric
interfaces in Cu/Co multilayer. The thermal stability of the
multilayer improves due to reduced inter-diffusion.

\acknowledgements {We acknowledge DST, Government of India for
providing financial support to carry out NR experiments under its
scheme `Utilization of International Synchrotron Radiation and
Neutron Scattering facilities'. A part of this work was performed
under the Indo-Swiss Joint Research Programme with grant no.
INT/SWISS/JUAF(9)/2009. Thanks to S.\,Potdar for the help provided
in sample preparation.}


\begin{thebibliography}{10}%
\makeatletter
\providecommand \@ifxundefined [1]{%
 \ifx #1\undefined \expandafter \@firstoftwo
 \else \expandafter \@secondoftwo
\fi
}%
\providecommand \@ifnum [1]{%
 \ifnum #1\expandafter \@firstoftwo
 \else \expandafter \@secondoftwo
\fi
}%
\providecommand \enquote [1]{``#1''}%
\providecommand \bibnamefont  [1]{#1}%
\providecommand \bibfnamefont [1]{#1}%
\providecommand \citenamefont [1]{#1}%
\providecommand\href[0]{\@sanitize\@href}%
\providecommand\@href[1]{\endgroup\@@startlink{#1}\endgroup\@@href}%
\providecommand\@@href[1]{#1\@@endlink}%
\providecommand \@sanitize [0]{\begingroup\catcode`\&12\catcode`\#12\relax}%
\@ifxundefined \pdfoutput {\@firstoftwo}{%
 \@ifnum{\z@=\pdfoutput}{\@firstoftwo}{\@secondoftwo}%
}{%
 \providecommand\@@startlink[1]{\leavevmode}%
 \providecommand\@@endlink[0]{}%
}{%
 \providecommand\@@startlink[1]{%
  \leavevmode
  \pdfstartlink
   attr{/Border[0 0 1 ]/H/I/C[0 1 1]}%
   user{/Subtype/Link/A<</Type/Action/S/URI/URI(#1)>>}%
  \relax
 }%
 \providecommand\@@endlink[0]{\pdfendlink}%
}%
\providecommand \url  [0]{\begingroup\@sanitize \@url }%
\providecommand \@url [1]{\endgroup\@href {#1}{\urlprefix}}%
\providecommand \urlprefix [0]{URL }%
\providecommand \Eprint[0]{\href }%
\@ifxundefined \urlstyle {%
  \providecommand \doi [1]{doi:\discretionary{}{}{}#1}%
}{%
  \providecommand \doi [0]{doi:\discretionary{}{}{}\begingroup
  \urlstyle{rm}\Url }%
}%
\providecommand \doibase [0]{http://dx.doi.org/}%
\providecommand \Doi[1]{\href{\doibase#1}}%
\providecommand \selectlanguage [0]{\@gobble}%
\providecommand \bibinfo [0]{\@secondoftwo}%
\providecommand \bibfield [0]{\@secondoftwo}%
\providecommand \translation [1]{[#1]}%
\providecommand \BibitemOpen[0]{}%
\providecommand \bibitemStop [0]{}%
\providecommand \bibitemNoStop [0]{.\EOS\space}%
\providecommand \EOS [0]{\spacefactor3000\relax}%
\providecommand \BibitemShut [1]{\csname bibitem#1\endcsname}%
\bibitem{Wolf_Science2001}%
  \BibitemOpen
  \bibfield{author}{%
  \bibinfo {author} {\bibfnamefont{S.~A.}\ \bibnamefont{Wolf}}, \bibinfo
  {author} {\bibfnamefont{D.~D.}\ \bibnamefont{Awschalom}}, \bibinfo {author}
  {\bibfnamefont{R.~A.}\ \bibnamefont{Buhrman}}, \bibinfo {author}
  {\bibfnamefont{J.~M.}\ \bibnamefont{Daughton}}, \bibinfo {author}
  {\bibfnamefont{S.}~\bibnamefont{von Molnár}}, \bibinfo {author}
  {\bibfnamefont{M.~L.}\ \bibnamefont{Roukes}}, \bibinfo {author}
  {\bibfnamefont{A.~Y.}\ \bibnamefont{Chtchelkanova}},\ and\ \bibinfo {author}
  {\bibfnamefont{D.~M.}\ \bibnamefont{Treger}},\ }%
  \bibfield{journal}{%
  \Doi{10.1126/science.1065389}{\bibinfo {journal} {Science}}\ }%
  \textbf{\bibinfo {volume} {294}},\ \bibinfo {pages} {1488} (\bibinfo {year}
  {2001})\BibitemShut{NoStop}%
\bibitem{Prinz_Science98}%
  \BibitemOpen
  \bibfield{author}{%
  \bibinfo {author} {\bibfnamefont{G.~A.}\ \bibnamefont{Prinz}},\ }%
  \bibfield{journal}{%
  \Doi{10.1126/science.282.5394.1660}{\bibinfo {journal} {Science}}\ }%
  \textbf{\bibinfo {volume} {282}},\ \bibinfo {pages} {1660} (\bibinfo {year}
  {1998})\BibitemShut{NoStop}%
\bibitem{carcia_JAP1988}%
  \BibitemOpen
  \bibfield{author}{%
  \bibinfo {author} {\bibfnamefont{P.~F.}\ \bibnamefont{Carcia}},\ }%
  \bibfield{journal}{%
  \Doi{10.1063/1.340404}{\bibinfo {journal} {J. Appl. Phys.}}\ }%
  \textbf{\bibinfo {volume} {63}},\ \bibinfo {pages} {5066} (\bibinfo {year}
  {1988})\BibitemShut{NoStop}%
\bibitem{Parkin_PRL1990}%
  \BibitemOpen
  \bibfield{author}{%
  \bibinfo {author} {\bibfnamefont{S.~S.~P.}\ \bibnamefont{Parkin}}, \bibinfo
  {author} {\bibfnamefont{N.}~\bibnamefont{More}},\ and\ \bibinfo {author}
  {\bibfnamefont{K.~P.}\ \bibnamefont{Roche}},\ }%
  \bibfield{journal}{%
  \Doi{10.1103/PhysRevLett.64.2304}{\bibinfo {journal} {Phys. Rev. Lett.}}\ }%
  \textbf{\bibinfo {volume} {64}},\ \bibinfo {pages} {2304} (\bibinfo {year}
  {1990})\BibitemShut{NoStop}%
\bibitem{Parkin_PRL1991}%
  \BibitemOpen
  \bibfield{author}{%
  \bibinfo {author} {\bibfnamefont{S.~S.~P.}\ \bibnamefont{Parkin}}, \bibinfo
  {author} {\bibfnamefont{R.}~\bibnamefont{Bhadra}},\ and\ \bibinfo {author}
  {\bibfnamefont{K.~P.}\ \bibnamefont{Roche}},\ }%
  \bibfield{journal}{%
  \Doi{10.1103/PhysRevLett.66.2152}{\bibinfo {journal} {Phys. Rev. Lett.}}\ }%
  \textbf{\bibinfo {volume} {66}},\ \bibinfo {pages} {2152} (\bibinfo {year}
  {1991})\BibitemShut{NoStop}%
\bibitem{Schmidt_PRB1999}%
  \BibitemOpen
  \bibfield{author}{%
  \bibinfo {author} {\bibfnamefont{C.~M.}\ \bibnamefont{Schmidt}}, \bibinfo
  {author} {\bibfnamefont{D.~E.}\ \bibnamefont{B\"urgler}}, \bibinfo {author}
  {\bibfnamefont{D.~M.}\ \bibnamefont{Schaller}}, \bibinfo {author}
  {\bibfnamefont{F.}~\bibnamefont{Meisinger}},\ and\ \bibinfo {author}
  {\bibfnamefont{H.-J.}\ \bibnamefont{G\"untherodt}},\ }%
  \bibfield{journal}{%
  \Doi{10.1103/PhysRevB.60.4158}{\bibinfo {journal} {Phys. Rev. B}}\ }%
  \textbf{\bibinfo {volume} {60}},\ \bibinfo {pages} {4158} (\bibinfo {year}
  {1999})\BibitemShut{NoStop}%
\bibitem{Parkin_PRL1993}%
  \BibitemOpen
  \bibfield{author}{%
  \bibinfo {author} {\bibfnamefont{S.~S.~P.}\ \bibnamefont{Parkin}},\ }%
  \bibfield{journal}{%
  \Doi{10.1103/PhysRevLett.71.1641}{\bibinfo {journal} {Phys. Rev. Lett.}}\ }%
  \textbf{\bibinfo {volume} {71}},\ \bibinfo {pages} {1641} (\bibinfo {year}
  {1993})\BibitemShut{NoStop}%
\bibitem{Zahn_PRL98}%
  \BibitemOpen
  \bibfield{author}{%
  \bibinfo {author} {\bibfnamefont{P.}~\bibnamefont{Zahn}}, \bibinfo {author}
  {\bibfnamefont{J.}~\bibnamefont{Binder}}, \bibinfo {author}
  {\bibfnamefont{I.}~\bibnamefont{Mertig}}, \bibinfo {author}
  {\bibfnamefont{R.}~\bibnamefont{Zeller}},\ and\ \bibinfo {author}
  {\bibfnamefont{P.~H.}\ \bibnamefont{Dederichs}},\ }%
  \bibfield{journal}{%
  \Doi{10.1103/PhysRevLett.80.4309}{\bibinfo {journal} {Phys. Rev. Lett.}}\ }%
  \textbf{\bibinfo {volume} {80}},\ \bibinfo {pages} {4309} (\bibinfo {year}
  {1998})\BibitemShut{NoStop}%
\bibitem{Kief_PRB93}%
  \BibitemOpen
  \bibfield{author}{%
  \bibinfo {author} {\bibfnamefont{M.~T.}\ \bibnamefont{Kief}}\ and\ \bibinfo
  {author} {\bibfnamefont{W.~F.}\ \bibnamefont{Egelhoff}},\ }%
  \bibfield{journal}{%
  \Doi{10.1103/PhysRevB.47.10785}{\bibinfo {journal} {Phys. Rev. B}}\ }%
  \textbf{\bibinfo {volume} {47}},\ \bibinfo {pages} {10785} (\bibinfo {year}
  {1993})\BibitemShut{NoStop}%
\bibitem{Baibich:PRL88}%
  \BibitemOpen
  \bibfield{author}{%
  \bibinfo {author} {\bibfnamefont{M.~N.}\ \bibnamefont{Baibich}}, \bibinfo
  {author} {\bibfnamefont{J.~M.}\ \bibnamefont{Broto}}, \bibinfo {author}
  {\bibfnamefont{A.}~\bibnamefont{Fert}}, \bibinfo {author}
  {\bibfnamefont{F.~N.}\ \bibnamefont{Van~Dau}}, \bibinfo {author}
  {\bibfnamefont{F.}~\bibnamefont{Petroff}}, \bibinfo {author}
  {\bibfnamefont{P.}~\bibnamefont{Etienne}}, \bibinfo {author}
  {\bibfnamefont{G.}~\bibnamefont{Creuzet}}, \bibinfo {author}
  {\bibfnamefont{A.}~\bibnamefont{Friederich}},\ and\ \bibinfo {author}
  {\bibfnamefont{J.}~\bibnamefont{Chazelas}},\ }%
  \bibfield{journal}{%
  \Doi{10.1103/PhysRevLett.61.2472}{\bibinfo {journal} {Phys. Rev. Lett.}}\ }%
  \textbf{\bibinfo {volume} {61}},\ \bibinfo {pages} {2472} (\bibinfo {year}
  {1988})\BibitemShut{NoStop}%
\bibitem{Fullerton:PRL92}%
  \BibitemOpen
  \bibfield{author}{%
  \bibinfo {author} {\bibfnamefont{E.~E.}\ \bibnamefont{Fullerton}}, \bibinfo
  {author} {\bibfnamefont{D.~M.}\ \bibnamefont{Kelly}}, \bibinfo {author}
  {\bibfnamefont{J.}~\bibnamefont{Guimpel}}, \bibinfo {author}
  {\bibfnamefont{I.~K.}\ \bibnamefont{Schuller}},\ and\ \bibinfo {author}
  {\bibfnamefont{Y.}~\bibnamefont{Bruynseraede}},\ }%
  \bibfield{journal}{%
  \Doi{10.1103/PhysRevLett.68.859}{\bibinfo {journal} {Phys. Rev. Lett.}}\ }%
  \textbf{\bibinfo {volume} {68}},\ \bibinfo {pages} {859} (\bibinfo {year}
  {1992})\BibitemShut{NoStop}%
\bibitem{Hall:PRB93}%
  \BibitemOpen
  \bibfield{author}{%
  \bibinfo {author} {\bibfnamefont{M.~J.}\ \bibnamefont{Hall}}, \bibinfo
  {author} {\bibfnamefont{B.~J.}\ \bibnamefont{Hickey}}, \bibinfo {author}
  {\bibfnamefont{M.~A.}\ \bibnamefont{Howson}}, \bibinfo {author}
  {\bibfnamefont{M.~J.}\ \bibnamefont{Walker}}, \bibinfo {author}
  {\bibfnamefont{J.}~\bibnamefont{Xu}}, \bibinfo {author}
  {\bibfnamefont{D.}~\bibnamefont{Greig}},\ and\ \bibinfo {author}
  {\bibfnamefont{N.}~\bibnamefont{Wiser}},\ }%
  \bibfield{journal}{%
  \Doi{10.1103/PhysRevB.47.12785}{\bibinfo {journal} {Phys. Rev. B}}\ }%
  \textbf{\bibinfo {volume} {47}},\ \bibinfo {pages} {12785} (\bibinfo {year}
  {1993})\BibitemShut{NoStop}%
\bibitem{egelhoff_JVT1989}%
  \BibitemOpen
  \bibfield{author}{%
  \bibinfo {author} {\bibfnamefont{J.}~\bibnamefont{W.~F.~Egelhoff}}\ and\
  \bibinfo {author} {\bibfnamefont{D.~A.}\ \bibnamefont{Steigerwald}},\ }%
  \bibfield{journal}{%
  \Doi{10.1116/1.575951}{\bibinfo {journal} {J. Vac. Sci. Techn. A}}\ }%
  \textbf{\bibinfo {volume} {7}},\ \bibinfo {pages} {2167} (\bibinfo {year}
  {1989})\BibitemShut{NoStop}%
\bibitem{Foiles_PRB86}%
  \BibitemOpen
  \bibfield{author}{%
  \bibinfo {author} {\bibfnamefont{S.~M.}\ \bibnamefont{Foiles}}, \bibinfo
  {author} {\bibfnamefont{M.~I.}\ \bibnamefont{Baskes}},\ and\ \bibinfo
  {author} {\bibfnamefont{M.~S.}\ \bibnamefont{Daw}},\ }%
  \bibfield{journal}{%
  \Doi{10.1103/PhysRevB.33.7983}{\bibinfo {journal} {Phys. Rev. B}}\ }%
  \textbf{\bibinfo {volume} {33}},\ \bibinfo {pages} {7983} (\bibinfo {year}
  {1986})\BibitemShut{NoStop}%
\bibitem{Tyson_SS77}%
  \BibitemOpen
  \bibfield{author}{%
  \bibinfo {author} {\bibfnamefont{W.}~\bibnamefont{Tyson}}\ and\ \bibinfo
  {author} {\bibfnamefont{W.}~\bibnamefont{Miller}},\ }%
  \bibfield{journal}{%
  \bibinfo {journal} {Surface Science}\ }%
  \textbf{\bibinfo {volume} {62}},\ \bibinfo {pages} {267 } (\bibinfo {year}
  {1977})\BibitemShut{NoStop}%
\bibitem{Timothy:JAP96}%
  \BibitemOpen
  \bibfield{author}{%
  \bibinfo {author} {\bibfnamefont{T.~J.}\ \bibnamefont{Minvielle}}, \bibinfo
  {author} {\bibfnamefont{R.~L.}\ \bibnamefont{White}},\ and\ \bibinfo {author}
  {\bibfnamefont{R.~J.}\ \bibnamefont{Wilson}},\ }%
  \bibfield{journal}{%
  \Doi{10.1063/1.361320}{\bibinfo {journal} {J. Appl. Phys.}}\ }%
  \textbf{\bibinfo {volume} {79}},\ \bibinfo {pages} {5116} (\bibinfo {year}
  {1996})\BibitemShut{NoStop}%
\bibitem{Buckley}%
  \BibitemOpen
  \bibfield{author}{%
  \bibinfo {author} {\bibfnamefont{H.~E.}\ \bibnamefont{Buckley}},\ }%
  \emph{\bibinfo {title} {Crystal Growth}}\ (\bibinfo {publisher} {Wiley, New
  York},\ \bibinfo {year} {1950})\BibitemShut{NoStop}%
\bibitem{Copel_PRL89}%
  \BibitemOpen
  \bibfield{author}{%
  \bibinfo {author} {\bibfnamefont{M.}~\bibnamefont{Copel}}, \bibinfo {author}
  {\bibfnamefont{M.~C.}\ \bibnamefont{Reuter}}, \bibinfo {author}
  {\bibfnamefont{E.}~\bibnamefont{Kaxiras}},\ and\ \bibinfo {author}
  {\bibfnamefont{R.~M.}\ \bibnamefont{Tromp}},\ }%
  \bibfield{journal}{%
  \Doi{10.1103/PhysRevLett.63.632}{\bibinfo {journal} {Phys. Rev. Lett.}}\ }%
  \textbf{\bibinfo {volume} {63}},\ \bibinfo {pages} {632} (\bibinfo {year}
  {1989})\BibitemShut{NoStop}%
\bibitem{Egelhoff_JAP1996}%
  \BibitemOpen
  \bibfield{author}{%
  \bibinfo {author} {\bibfnamefont{J.}~\bibnamefont{W.~F.~Egelhoff}}, \bibinfo
  {author} {\bibfnamefont{P.~J.}\ \bibnamefont{Chen}}, \bibinfo {author}
  {\bibfnamefont{C.~J.}\ \bibnamefont{Powell}}, \bibinfo {author}
  {\bibfnamefont{M.~D.}\ \bibnamefont{Stiles}},\ and\ \bibinfo {author}
  {\bibfnamefont{R.~D.}\ \bibnamefont{McMichael}},\ }%
  \bibfield{journal}{%
  \Doi{10.1063/1.362659}{\bibinfo {journal} {J. Appl. Phys.}}\ }%
  \textbf{\bibinfo {volume} {79}},\ \bibinfo {pages} {2491} (\bibinfo {year}
  {1996})\BibitemShut{NoStop}%
\bibitem{Egelhoff_JAP96_Pb_Au}%
  \BibitemOpen
  \bibfield{author}{%
  \bibinfo {author} {\bibfnamefont{J.}~\bibnamefont{W.~F.~Egelhoff}}, \bibinfo
  {author} {\bibfnamefont{P.~J.}\ \bibnamefont{Chen}}, \bibinfo {author}
  {\bibfnamefont{C.~J.}\ \bibnamefont{Powell}}, \bibinfo {author}
  {\bibfnamefont{M.~D.}\ \bibnamefont{Stiles}}, \bibinfo {author}
  {\bibfnamefont{R.~D.}\ \bibnamefont{McMichael}}, \bibinfo {author}
  {\bibfnamefont{C.-L.}\ \bibnamefont{Lin}}, \bibinfo {author}
  {\bibfnamefont{J.~M.}\ \bibnamefont{Sivertsen}}, \bibinfo {author}
  {\bibfnamefont{J.~H.}\ \bibnamefont{Judy}}, \bibinfo {author}
  {\bibfnamefont{K.}~\bibnamefont{Takano}},\ and\ \bibinfo {author}
  {\bibfnamefont{A.~E.}\ \bibnamefont{Berkowitz}},\ }%
  \bibfield{journal}{%
  \Doi{10.1063/1.363460}{\bibinfo {journal} {J. Appl. Phys.}}\ }%
  \textbf{\bibinfo {volume} {80}},\ \bibinfo {pages} {5183} (\bibinfo {year}
  {1996})\BibitemShut{NoStop}%
\bibitem{Egelhof_JAP97_O}%
  \BibitemOpen
  \bibfield{author}{%
  \bibinfo {author} {\bibfnamefont{J.}~\bibnamefont{W.~F.~Egelhoff}}, \bibinfo
  {author} {\bibfnamefont{P.~J.}\ \bibnamefont{Chen}}, \bibinfo {author}
  {\bibfnamefont{C.~J.}\ \bibnamefont{Powell}}, \bibinfo {author}
  {\bibfnamefont{M.~D.}\ \bibnamefont{Stiles}}, \bibinfo {author}
  {\bibfnamefont{R.~D.}\ \bibnamefont{McMichael}}, \bibinfo {author}
  {\bibfnamefont{J.~H.}\ \bibnamefont{Judy}}, \bibinfo {author}
  {\bibfnamefont{K.}~\bibnamefont{Takano}},\ and\ \bibinfo {author}
  {\bibfnamefont{A.~E.}\ \bibnamefont{Berkowitz}},\ }%
  \bibfield{journal}{%
  \Doi{10.1063/1.365620}{\bibinfo {journal} {J. Appl. Phys.}}\ }%
  \textbf{\bibinfo {volume} {82}},\ \bibinfo {pages} {6142} (\bibinfo {year}
  {1997})\BibitemShut{NoStop}%
\bibitem{osten_APL92}%
  \BibitemOpen
  \bibfield{author}{%
  \bibinfo {author} {\bibfnamefont{H.~J.}\ \bibnamefont{Osten}}, \bibinfo
  {author} {\bibfnamefont{J.}~\bibnamefont{Klatt}}, \bibinfo {author}
  {\bibfnamefont{G.}~\bibnamefont{Lippert}}, \bibinfo {author}
  {\bibfnamefont{E.}~\bibnamefont{Bugiel}},\ and\ \bibinfo {author}
  {\bibfnamefont{S.}~\bibnamefont{Hinrich}},\ }%
  \bibfield{journal}{%
  \Doi{10.1063/1.106926}{\bibinfo {journal} {Appl. Phys. Lett.}}\ }%
  \textbf{\bibinfo {volume} {60}},\ \bibinfo {pages} {2522} (\bibinfo {year}
  {1992})\BibitemShut{NoStop}%
\bibitem{Camarero_PRL94}%
  \BibitemOpen
  \bibfield{author}{%
  \bibinfo {author} {\bibfnamefont{J.}~\bibnamefont{Camarero}}, \bibinfo
  {author} {\bibfnamefont{L.}~\bibnamefont{Spendeler}}, \bibinfo {author}
  {\bibfnamefont{G.}~\bibnamefont{Schmidt}}, \bibinfo {author}
  {\bibfnamefont{K.}~\bibnamefont{Heinz}}, \bibinfo {author}
  {\bibfnamefont{J.~J.}\ \bibnamefont{de~Miguel}},\ and\ \bibinfo {author}
  {\bibfnamefont{R.}~\bibnamefont{Miranda}},\ }%
  \bibfield{journal}{%
  \Doi{10.1103/PhysRevLett.73.2448}{\bibinfo {journal} {Phys. Rev. Lett.}}\ }%
  \textbf{\bibinfo {volume} {73}},\ \bibinfo {pages} {2448} (\bibinfo {year}
  {1994})\BibitemShut{NoStop}%
\bibitem{Camarero_PRL96}%
  \BibitemOpen
  \bibfield{author}{%
  \bibinfo {author} {\bibfnamefont{J.}~\bibnamefont{Camarero}}, \bibinfo
  {author} {\bibfnamefont{T.}~\bibnamefont{Graf}}, \bibinfo {author}
  {\bibfnamefont{J.~J.}\ \bibnamefont{de~Miguel}}, \bibinfo {author}
  {\bibfnamefont{R.}~\bibnamefont{Miranda}}, \bibinfo {author}
  {\bibfnamefont{W.}~\bibnamefont{Kuch}}, \bibinfo {author}
  {\bibfnamefont{M.}~\bibnamefont{Zharnikov}}, \bibinfo {author}
  {\bibfnamefont{A.}~\bibnamefont{Dittschar}}, \bibinfo {author}
  {\bibfnamefont{C.~M.}\ \bibnamefont{Schneider}},\ and\ \bibinfo {author}
  {\bibfnamefont{J.}~\bibnamefont{Kirschner}},\ }%
  \bibfield{journal}{%
  \Doi{10.1103/PhysRevLett.76.4428}{\bibinfo {journal} {Phys. Rev. Lett.}}\ }%
  \textbf{\bibinfo {volume} {76}},\ \bibinfo {pages} {4428} (\bibinfo {year}
  {1996})\BibitemShut{NoStop}%
\bibitem{Chopra_PRB97}%
  \BibitemOpen
  \bibfield{author}{%
  \bibinfo {author} {\bibfnamefont{H.~D.}\ \bibnamefont{Chopra}}, \bibinfo
  {author} {\bibfnamefont{B.~J.}\ \bibnamefont{Hockey}}, \bibinfo {author}
  {\bibfnamefont{P.~J.}\ \bibnamefont{Chen}}, \bibinfo {author}
  {\bibfnamefont{W.~F.}\ \bibnamefont{Egelhoff}}, \bibinfo {author}
  {\bibfnamefont{M.}~\bibnamefont{Wuttig}},\ and\ \bibinfo {author}
  {\bibfnamefont{S.~Z.}\ \bibnamefont{Hua}},\ }%
  \bibfield{journal}{%
  \Doi{10.1103/PhysRevB.55.8390}{\bibinfo {journal} {Phys. Rev. B}}\ }%
  \textbf{\bibinfo {volume} {55}},\ \bibinfo {pages} {8390} (\bibinfo {year}
  {1997})\BibitemShut{NoStop}%
\bibitem{Tolkes_PRL98}%
  \BibitemOpen
  \bibfield{author}{%
  \bibinfo {author} {\bibfnamefont{C.}~\bibnamefont{T\"olkes}}, \bibinfo
  {author} {\bibfnamefont{R.}~\bibnamefont{Struck}}, \bibinfo {author}
  {\bibfnamefont{R.}~\bibnamefont{David}}, \bibinfo {author}
  {\bibfnamefont{P.}~\bibnamefont{Zeppenfeld}},\ and\ \bibinfo {author}
  {\bibfnamefont{G.}~\bibnamefont{Comsa}},\ }%
  \bibfield{journal}{%
  \Doi{10.1103/PhysRevLett.80.2877}{\bibinfo {journal} {Phys. Rev. Lett.}}\ }%
  \textbf{\bibinfo {volume} {80}},\ \bibinfo {pages} {2877} (\bibinfo {year}
  {1998})\BibitemShut{NoStop}%
\bibitem{Wen_PRB2007}%
  \BibitemOpen
  \bibfield{author}{%
  \bibinfo {author} {\bibfnamefont{H.}~\bibnamefont{Wen}}, \bibinfo {author}
  {\bibfnamefont{M.}~\bibnamefont{Neurock}},\ and\ \bibinfo {author}
  {\bibfnamefont{H.~N.~G.}\ \bibnamefont{Wadley}},\ }%
  \bibfield{journal}{%
  \Doi{10.1103/PhysRevB.75.085403}{\bibinfo {journal} {Phys. Rev. B}}\ }%
  \textbf{\bibinfo {volume} {75}},\ \bibinfo {pages} {085403} (\bibinfo {year}
  {2007})\BibitemShut{NoStop}%
\bibitem{Larson_PRB2003}%
  \BibitemOpen
  \bibfield{author}{%
  \bibinfo {author} {\bibfnamefont{D.~J.}\ \bibnamefont{Larson}}, \bibinfo
  {author} {\bibfnamefont{A.~K.}\ \bibnamefont{Petford-Long}}, \bibinfo
  {author} {\bibfnamefont{A.}~\bibnamefont{Cerezo}}, \bibinfo {author}
  {\bibfnamefont{S.~P.}\ \bibnamefont{Bozeman}}, \bibinfo {author}
  {\bibfnamefont{A.}~\bibnamefont{Morrone}}, \bibinfo {author}
  {\bibfnamefont{Y.~Q.}\ \bibnamefont{Ma}}, \bibinfo {author}
  {\bibfnamefont{A.}~\bibnamefont{Georgalakis}},\ and\ \bibinfo {author}
  {\bibfnamefont{P.~H.}\ \bibnamefont{Clifton}},\ }%
  \bibfield{journal}{%
  \Doi{10.1103/PhysRevB.67.144420}{\bibinfo {journal} {Phys. Rev. B}}\ }%
  \textbf{\bibinfo {volume} {67}},\ \bibinfo {pages} {144420} (\bibinfo {year}
  {2003})\BibitemShut{NoStop}%
\bibitem{Hoegen_PRL91}%
  \BibitemOpen
  \bibfield{author}{%
  \bibinfo {author} {\bibfnamefont{M.}~\bibnamefont{Horn-von Hoegen}}, \bibinfo
  {author} {\bibfnamefont{F.~K.}\ \bibnamefont{LeGoues}}, \bibinfo {author}
  {\bibfnamefont{M.}~\bibnamefont{Copel}}, \bibinfo {author}
  {\bibfnamefont{M.~C.}\ \bibnamefont{Reuter}},\ and\ \bibinfo {author}
  {\bibfnamefont{R.~M.}\ \bibnamefont{Tromp}},\ }%
  \bibfield{journal}{%
  \Doi{10.1103/PhysRevLett.67.1130}{\bibinfo {journal} {Phys. Rev. Lett.}}\ }%
  \textbf{\bibinfo {volume} {67}},\ \bibinfo {pages} {1130} (\bibinfo {year}
  {1991})\BibitemShut{NoStop}%
\bibitem{Barabasi_PRL1993}%
  \BibitemOpen
  \bibfield{author}{%
  \bibinfo {author} {\bibfnamefont{A.-L.}\ \bibnamefont{Barab\'asi}},\ }%
  \bibfield{journal}{%
  \Doi{10.1103/PhysRevLett.70.4102}{\bibinfo {journal} {Phys. Rev. Lett.}}\ }%
  \textbf{\bibinfo {volume} {70}},\ \bibinfo {pages} {4102} (\bibinfo {year}
  {1993})\BibitemShut{NoStop}%
\bibitem{Zhang_PRL94}%
  \BibitemOpen
  \bibfield{author}{%
  \bibinfo {author} {\bibfnamefont{Z.}~\bibnamefont{Zhang}}\ and\ \bibinfo
  {author} {\bibfnamefont{M.~G.}\ \bibnamefont{Lagally}},\ }%
  \bibfield{journal}{%
  \Doi{10.1103/PhysRevLett.72.693}{\bibinfo {journal} {Phys. Rev. Lett.}}\ }%
  \textbf{\bibinfo {volume} {72}},\ \bibinfo {pages} {693} (\bibinfo {year}
  {1994})\BibitemShut{NoStop}%
\bibitem{yang_JAP2001}%
  \BibitemOpen
  \bibfield{author}{%
  \bibinfo {author} {\bibfnamefont{D.~X.}\ \bibnamefont{Yang}}, \bibinfo
  {author} {\bibfnamefont{B.}~\bibnamefont{Shashishekar}}, \bibinfo {author}
  {\bibfnamefont{H.~D.}\ \bibnamefont{Chopra}}, \bibinfo {author}
  {\bibfnamefont{P.~J.}\ \bibnamefont{Chen}},\ and\ \bibinfo {author}
  {\bibfnamefont{J.}~\bibnamefont{W.~F.~Egelhoff}},\ }%
  \bibfield{journal}{%
  \Doi{10.1063/1.1359225}{\bibinfo {journal} {J. Appl. Phys.}}\ }%
  \textbf{\bibinfo {volume} {89}},\ \bibinfo {pages} {7121} (\bibinfo {year}
  {2001})\BibitemShut{NoStop}%
\bibitem{Chopra:PRB2002}%
  \BibitemOpen
  \bibfield{author}{%
  \bibinfo {author} {\bibfnamefont{H.~D.}\ \bibnamefont{Chopra}}, \bibinfo
  {author} {\bibfnamefont{D.~X.}\ \bibnamefont{Yang}}, \bibinfo {author}
  {\bibfnamefont{P.~J.}\ \bibnamefont{Chen}},\ and\ \bibinfo {author}
  {\bibfnamefont{W.~F.}\ \bibnamefont{Egelhoff}},\ }%
  \bibfield{journal}{%
  \Doi{10.1103/PhysRevB.65.094433}{\bibinfo {journal} {Phys. Rev. B}}\ }%
  \textbf{\bibinfo {volume} {65}},\ \bibinfo {pages} {094433} (\bibinfo {year}
  {2002})\BibitemShut{NoStop}%
\bibitem{Gomez:PRB01}%
  \BibitemOpen
  \bibfield{author}{%
  \bibinfo {author} {\bibfnamefont{L.}~\bibnamefont{G\'omez}}\ and\ \bibinfo
  {author} {\bibfnamefont{J.}~\bibnamefont{Ferr\'on}},\ }%
  \bibfield{journal}{%
  \Doi{10.1103/PhysRevB.64.033409}{\bibinfo {journal} {Phys. Rev. B}}\ }%
  \textbf{\bibinfo {volume} {64}},\ \bibinfo {pages} {033409} (\bibinfo {year}
  {2001})\BibitemShut{NoStop}%
\bibitem{kim:JAP2009}%
  \BibitemOpen
  \bibfield{author}{%
  \bibinfo {author} {\bibfnamefont{B.-H.}\ \bibnamefont{Kim}}\ and\ \bibinfo
  {author} {\bibfnamefont{Y.-C.}\ \bibnamefont{Chung}},\ }%
  \bibfield{journal}{%
  \Doi{10.1063/1.3194309}{\bibinfo {journal} {J. Appl. Phys.s}}\ }%
  \textbf{\bibinfo {volume} {106}},\ \bibinfo {eid} {044304} (\bibinfo {year}
  {2009})\BibitemShut{NoStop}%
\bibitem{an_JAP2006}%
  \BibitemOpen
  \bibfield{author}{%
  \bibinfo {author} {\bibfnamefont{Y.}~\bibnamefont{An}}, \bibinfo {author}
  {\bibfnamefont{H.}~\bibnamefont{Zhang}}, \bibinfo {author}
  {\bibfnamefont{B.}~\bibnamefont{Dai}}, \bibinfo {author}
  {\bibfnamefont{Z.}~\bibnamefont{Mai}}, \bibinfo {author}
  {\bibfnamefont{J.}~\bibnamefont{Cai}},\ and\ \bibinfo {author}
  {\bibfnamefont{Z.}~\bibnamefont{Wu}},\ }%
  \bibfield{journal}{%
  \Doi{10.1063/1.2214369}{\bibinfo {journal} {J. Appl. Phys.}}\ }%
  \textbf{\bibinfo {volume} {100}},\ \bibinfo {eid} {023516} (\bibinfo {year}
  {2006})\BibitemShut{NoStop}%
\bibitem{Peterson:PhysicaB2003}%
  \BibitemOpen
  \bibfield{author}{%
  \bibinfo {author} {\bibfnamefont{B.~L.}\ \bibnamefont{Peterson}}, \bibinfo
  {author} {\bibfnamefont{R.~L.}\ \bibnamefont{White}},\ and\ \bibinfo {author}
  {\bibfnamefont{B.~M.}\ \bibnamefont{Clemens}},\ }%
  \bibfield{journal}{%
  \Doi{DOI: 10.1016/S0921-4526(03)00285-0}{\bibinfo {journal} {Physica B: Cond.
  Mat.}}\ }%
  \textbf{\bibinfo {volume} {336}},\ \bibinfo {pages} {157 } (\bibinfo {year}
  {2003})\BibitemShut{NoStop}%
\bibitem{Zou:PRB2001}%
  \BibitemOpen
  \bibfield{author}{%
  \bibinfo {author} {\bibfnamefont{W.}~\bibnamefont{Zou}}, \bibinfo {author}
  {\bibfnamefont{H.~N.~G.}\ \bibnamefont{Wadley}}, \bibinfo {author}
  {\bibfnamefont{X.~W.}\ \bibnamefont{Zhou}}, \bibinfo {author}
  {\bibfnamefont{R.~A.}\ \bibnamefont{Johnson}},\ and\ \bibinfo {author}
  {\bibfnamefont{D.}~\bibnamefont{Brownell}},\ }%
  \bibfield{journal}{%
  \Doi{10.1103/PhysRevB.64.174418}{\bibinfo {journal} {Phys. Rev. B}}\ }%
  \textbf{\bibinfo {volume} {64}},\ \bibinfo {pages} {174418} (\bibinfo {year}
  {2001})\BibitemShut{NoStop}%
\bibitem{TiNi_arXiv}%
  \BibitemOpen
  \bibfield{author}{%
  \bibinfo {author} {\bibfnamefont{M.}~\bibnamefont{Gupta}}, \bibinfo {author}
  {\bibfnamefont{S.~M.}\ \bibnamefont{Amir}}, \bibinfo {author}
  {\bibfnamefont{A.}~\bibnamefont{Gupta}},\ and\ \bibinfo {author}
  {\bibfnamefont{J.}~\bibnamefont{Stahn}},\ }%
  \bibfield{journal}{%
  \bibinfo {journal} {arXiv:1102.0688v1 [cond-mat.mtrl-sci]}}%
   (\bibinfo {year} {2011})\BibitemShut{NoStop}%
\bibitem{pi-pf_PRL_2002}%
  \BibitemOpen
  \bibfield{author}{%
  \bibinfo {author} {\bibfnamefont{V.}~\bibnamefont{Lauter-Pasyuk}}, \bibinfo
  {author} {\bibfnamefont{H.~J.}\ \bibnamefont{Lauter}}, \bibinfo {author}
  {\bibfnamefont{B.~P.}\ \bibnamefont{Toperverg}}, \bibinfo {author}
  {\bibfnamefont{L.}~\bibnamefont{Romashev}},\ and\ \bibinfo {author}
  {\bibfnamefont{V.}~\bibnamefont{Ustinov}},\ }%
  \bibfield{journal}{%
  \Doi{10.1103/PhysRevLett.89.167203}{\bibinfo {journal} {Phys. Rev. Lett.}}\
  }%
  \textbf{\bibinfo {volume} {89}},\ \bibinfo {pages} {167203} (\bibinfo {year}
  {2002})\BibitemShut{NoStop}%
\bibitem{Langridge:PRL2000}%
  \BibitemOpen
  \bibfield{author}{%
  \bibinfo {author} {\bibfnamefont{S.}~\bibnamefont{Langridge}}, \bibinfo
  {author} {\bibfnamefont{J.}~\bibnamefont{Schmalian}}, \bibinfo {author}
  {\bibfnamefont{C.~H.}\ \bibnamefont{Marrows}}, \bibinfo {author}
  {\bibfnamefont{D.~T.}\ \bibnamefont{Dekadjevi}},\ and\ \bibinfo {author}
  {\bibfnamefont{B.~J.}\ \bibnamefont{Hickey}},\ }%
  \bibfield{journal}{%
  \bibinfo {journal} {Phys. Rev. Lett.}\ }%
  \textbf{\bibinfo {volume} {85}},\ \bibinfo {pages} {4964} (\bibinfo {year}
  {2000})\BibitemShut{NoStop}%
\bibitem{Parratt32}%
  \BibitemOpen
  \bibfield{author}{%
  \bibinfo {author} {\bibfnamefont{C.}~\bibnamefont{Braun}},\ }%
  \emph{\bibinfo {title} {Parratt32- The Reflectivity Tool}}\ (\bibinfo
  {publisher} {HMI Berlin},\ \bibinfo {year} {1997-99})\BibitemShut{NoStop}%
\bibitem{MGupta_PRB2004}%
  \BibitemOpen
  \bibfield{author}{%
  \bibinfo {author} {\bibfnamefont{M.}~\bibnamefont{Gupta}}, \bibinfo {author}
  {\bibfnamefont{A.}~\bibnamefont{Gupta}}, \bibinfo {author}
  {\bibfnamefont{J.}~\bibnamefont{Stahn}}, \bibinfo {author}
  {\bibfnamefont{M.}~\bibnamefont{Horisberger}}, \bibinfo {author}
  {\bibfnamefont{T.}~\bibnamefont{Gutberlet}},\ and\ \bibinfo {author}
  {\bibfnamefont{P.}~\bibnamefont{Allenspach}},\ }%
  \bibfield{journal}{%
  \Doi{10.1103/PhysRevB.70.184206}{\bibinfo {journal} {Phys. Rev. B}}\ }%
  \textbf{\bibinfo {volume} {70}},\ \bibinfo {pages} {184206} (\bibinfo {year}
  {2004})\BibitemShut{NoStop}%
\bibitem{rosenblum_APL1980}%
  \BibitemOpen
  \bibfield{author}{%
  \bibinfo {author} {\bibfnamefont{M.~P.}\ \bibnamefont{Rosenblum}}, \bibinfo
  {author} {\bibfnamefont{F.}~\bibnamefont{Spaepen}},\ and\ \bibinfo {author}
  {\bibfnamefont{D.}~\bibnamefont{Turnbull}},\ }%
  \bibfield{journal}{%
  \Doi{10.1063/1.91818}{\bibinfo {journal} {Appl. Phys. Lett.}}\ }%
  \textbf{\bibinfo {volume} {37}},\ \bibinfo {pages} {184} (\bibinfo {year}
  {1980})\BibitemShut{NoStop}%
\bibitem{Schmidt:AM:2008}%
  \BibitemOpen
  \bibfield{author}{%
  \bibinfo {author} {\bibfnamefont{H.}~\bibnamefont{Schmidt}}, \bibinfo
  {author} {\bibfnamefont{M.}~\bibnamefont{Gupta}}, \bibinfo {author}
  {\bibfnamefont{T.}~\bibnamefont{Gutberlet}}, \bibinfo {author}
  {\bibfnamefont{J.}~\bibnamefont{Stahn}},\ and\ \bibinfo {author}
  {\bibfnamefont{M.}~\bibnamefont{Bruns}},\ }%
  \bibfield{journal}{%
  \bibinfo {journal} {Acta Mater.}\ }%
  \textbf{\bibinfo {volume} {56}},\ \bibinfo {pages} {464} (\bibinfo {year}
  {2008})\BibitemShut{NoStop}%
\bibitem{Li:PRB01}%
  \BibitemOpen
  \bibfield{author}{%
  \bibinfo {author} {\bibfnamefont{Z.~F.}\ \bibnamefont{Li}}, \bibinfo {author}
  {\bibfnamefont{Q.}~\bibnamefont{Zhang}}, \bibinfo {author}
  {\bibfnamefont{D.~P.}\ \bibnamefont{Yu}}, \bibinfo {author}
  {\bibfnamefont{C.}~\bibnamefont{Lin}},\ and\ \bibinfo {author}
  {\bibfnamefont{B.~X.}\ \bibnamefont{Liu}},\ }%
  \bibfield{journal}{%
  \Doi{10.1103/PhysRevB.64.014102}{\bibinfo {journal} {Phys. Rev. B}}\ }%
  \textbf{\bibinfo {volume} {64}},\ \bibinfo {pages} {014102} (\bibinfo {year}
  {2001})\BibitemShut{NoStop}%
\end{thebibliography}

%

\end{document}